\documentclass{aastex63}
\usepackage{graphicx}
\usepackage{amsmath}
\usepackage{enumitem}

\def\Msolar{\ifmmode {\rm M_{\odot}}\else $\rm M_{\odot}$\fi}
\def\Mearth{\ifmmode {\rm M_{\oplus}}\else $\rm M_{\oplus}$\fi}
\def\Rearth{\ifmmode {\rm R_{\oplus}}\else $\rm R_{\oplus}$\fi}
\def\micron{\ifmmode {\mu{\rm{m}}}\else $\mu$m\fi}

\def\inc{\imath}

\def\mfit{{\cal M}}

\begin{document}

\title{Polarization of circumstellar debris disk light echoes}

\author[0000-0001-9181-1105]{Austin J. King}
\author[0000-0001-7558-343X]{Benjamin C. Bromley}
\author[0000-0002-9576-2539]{Preston W. Harris}
\affil{Department of Physics \& Astronomy, University of Utah, 
\\ 115 S 1400 E, Rm 201, Salt Lake City, UT 84112}
\email{austin.king@utah.edu}

\author[0000-0003-0214-609X]{Scott J. Kenyon}
\affil{Smithsonian Astrophysical Observatory,
60 Garden Street, Cambridge, MA 02138}

\begin{abstract}
Light echoes of debris disks around active stars can reveal disk structure and composition even when disks are not spatially resolved. Unfortunately, distinguishing reflected light from quiescent starlight and unexpected post-peak flare structure is challenging, especially for edge-on geometries where the time delay between observed flare photons and light scattered from the near side of the disk is short. Here, we take advantage of the fact that scattered light from a dusty disk is polarized, depending on the location of the scattering site and the orientation of the disk relative to a distant observer. Filtering reflected light into its polarized components allows echoes to stand out in predictable ways. We test this idea with a simple model for a disk around an active M dwarf. Our results demonstrate that the use of polarimetric data of flaring stars can significantly enhance echo signals relative to starlight and yield more robust and accurate fits to disk parameters compared to analyses based on the total intensity alone. 
\end{abstract}


\section{Introduction} 
\label{intro}

Around young stars, growing planets emerge in partnership with dust. Within a protoplanetary disk, planets form when small solids are concentrated into progressively larger ones \citep{safronov1969, lissauer1987, wetherill1989, wetherill1993}. Giant impacts between growing protoplanets and shattering collisions among smaller planetesimals produce copious amounts of dust. This dust orbits the central star within disks or narrow rings \citep{najita2022}. Detection of this dusty debris can potentially reveal how the process of planet formation unfolds even when the planets themselves are undetectable \citep[e.g.,][see also \citealt{wyatt2008} and \citealt{hughes2018} for reviews]{kenyon2002, najita2022}.

Historically, debris disks have been detected in one of two ways \citep[for example,][]{aumann1985, plets1999, su2006, macgregor2013, hughes2018}. Small dust grains orbiting the central star are heated and then re-radiate starlight at a temperature lower than the temperature of the stellar photosphere. This process generates an excess detectable with infrared (IR) photometry, as in the discovery of a ring of dust around Vega \citep{aumann1984}.  Since this pathbreaking effort, more than 1000 debris disks have been discovered from 
their IR excess emission \citep[e.g.,][]{bryden2009, chen2014, sibthorpe2018, hillenbrand2008}.

Grains are also efficient at scattering light from the central star. If the dusty disk or ring is large enough, imaging observations can detect it. \citet{smithterrile} revealed the edge-on disk around the nearby A-type star $\beta$~Pic with ground-based coronagraphic imaging. AU Mic, an active M dwarf \citep{kalas2004}, also has a well-studied resolved debris disk \citep[e.g.,][]{matthews2015, lawson2023}. Another nearby A star, Fomalhaut, hosts a resolved ring of dust \citep{kalas2005} along with fainter debris  disk components \citep[e.g.,][]{Fomalhaut, holland1998, gaspar2023}. Although scattered-light observations have revealed fewer debris disks than photometric observations, recent discoveries with high-contrast adaptive optics instruments demonstrate a wide variety of debris architectures \citep[e.g.,][]{esposito2020}. 

Direct imaging and IR photometry are limited in their ability to detect debris. With a typical diffraction limit of 0.05--0.1 arcsec at a wavelength of 3--4~$\mu$m on a 6--10-m telescope, direct imaging can only resolve disks or rings at distances $\gtrsim$ 1--2~au around nearby stars at distances $\lesssim$ 10--100~pc \citep{macintosh08, macintosh2015, marois2008}. For a nearby solar-type star, it is a challenge to detect grains with temperatures much greater than 1000--2000~K, which generally places the debris beyond 0.5--1.0~au \citep[see][for example]{kenyon2016}.
In both methods, debris disk detections around stars with distance $\gtrsim$ 100--200~pc are generally limited to the brightest and youngest systems \citep[e.g.,][]{hughes2018}.

Toward probing dust that extends to Kuiper-belt distances around its host star, we consider the detection of debris disks in scattered light without the benefit of direct imaging. Scattered light can, in principle, constitute a significant fraction of the observed flux from a star. For example, a hypothetical, optically thick disk that extends from the surface of a star to arbitrarily large radial distance produces an observed flux that is equal to that of the star itself \citep{kenyon1987, chiang97, chiang99}. Even an optically thin circumstellar ring with a radius of 100 AU and a lunar mass of micron-size grains intercepts a few percent of the stellar host's light. Unfortunately, in observed systems, typically with less micron-size dust, the ratio of scattered light to the host starlight is $\sim$0.1\%\ or less \citep[e.g.,][]{kalas2004}. Thus, without direct imaging, disentangling the scattered light from the starlight is a challenge.  

Two strategies for studying scattered light from unresolved debris disks have emerged. One approach for disentangling the scattered light of an unresolved debris disk from the radiation of its host star is with light echoes \citep{gaidos1994}. Some active stars produce strong flares, sending bursts of light into the circumstellar environment like a sonar ping. The use of faint light echoes from stellar outbursts has been explored theoretically for planet detection \citep{argyle1974, bromley1992, Mann, Sparks2018}. \citet{gaidos1994} recognized that debris disks are more promising candidates for light echoes because of the enhanced scattering efficiency of many small grains compared to a single planet. Individual dust grains in a debris disk will reflect starlight, filling in an observed post-flare lightcurve depending on their location on the disk, light-travel time, and the orientation of the observer. The resulting lightcurve profile has a structure that is sensitive to these parameters and can, in principle, be fit with models to extract information about the disk and the grains that compose it.

In practice, light echoes from debris disks are certain to be faint, typically well below the signal to noise achieved in lightcurves from NASA's Kepler \citep{borucki2010} or TESS \citep{ricker2015} missions, for example. Two search strategies can mitigate this problem. The first is to focus on the most active stars, such as cool red dwarf flare stars (spectral type dMe). These objects produce flares in optical and UV light that can exceed the quiescent starlight in these wavebands by a factor of several hundred in extreme cases \citep{bond1976, kowalski2010, hawley2014, chang2018} to as much as $10^4$ in stars at the M--L spectral type boundary \citep{schmidt2014, schmidt2016}. In some cases, the flare rate is rapid, a few to tens of flares per day \citep{pettersen1986, hawley2014, balona2015, davenport2016, yang2017, pietras2022}. Frequent, powerful flares also tend to occur on stars that are younger, $<50$~Myr \citep[e.g.][]{feinstein2020}, at ages when debris disks are more common \citep[e.g.,][]{currie2008, pawellek21}.

The second strategy to overcome the faintness of light echos of debris disks is to combine lightcurves together, building up the signal-to-noise ratio \citep{bromley2021}. With extended monitoring campaigns like Kepler, we can select short-lived flares from individual stars and combine them; with up to hundreds of flares for some sources in the Kepler database, the average post-flare lightcurve can be fit with simple flare models of disk structure and grain properties. To achieve even better signal-to-noise, lightcurves from an ensemble of stars can be added together to constrain the average amount of dusty debris in the ensemble. \citet{bromley2021} combined over 34,000 flares in Kepler short-cadence data to limit the average mass of dust in the terrestrial zone to be less than about 0.1~lunar masses around late-type stars.

Our next approach for disentanglement of scattered light uses polarimetry: Dust, particularly micron-size grains, scatters light with a degree of polarization that depends on dust properties as well as the relative orientation of the star, the dust location, and the observer's perspective \citep{davis, draine2003, Das}. The underlying physics is that electromagnetic radiation from the host star causes electrons within dust grains to oscillate, launching light waves in response. From the perspective of a distant observer, the orientation of this scattered light is aligned with the projected oscillating motion of the electrons. For a disk observed face-on, electrons oscillate in directions mapped out by concentric circles about the host star; the polarization of the scattered light follows this pattern. For a disk observed edge on, scattered light coming from points at the projected extremities of the disk is polarized perpendicularly to the disk plane, while scattered light coming from the near (or far) edge of the disk along lines of sight close to the star is nearly unpolarized, like the starlight. Except for face-on disks where the polarized contributions from all points on the disk cancel out, the cumulative effect is that light scattered from a debris disk carries a net polarization, distinguishing it from the light of its stellar host. 

Numerous studies have revealed polarization in light scattered from dusty disks, providing a powerful diagnostic of disk structure and dust properties. Nearby resolved systems include AU~Mic \citep[e.g.,][HST]{graham2007}, $\beta$ Pic \citep{millar-blanchaer2015}, and HR~4796A \citep{milli2019, arriaga2020}. Polarization of scattered light from these stars and others \citep[e.g.,][]{asensio-torres2016, engler2017, bruzzone2020, esposito2020, crotts2021, perrot2023} provides important constraints on debris disk geometry, and the size and composition of the dusty debris. Other observations of polarized light have mapped dust around young stars \citep{wood1996, whitney1997, wood1998}, which showed a total polarization around $2\%$ but could reach as high as $11\%$, and even teased out photons from stars otherwise obscured by dust \citep[e.g.,][]{whitney1993}. Throughout, theoretical calculations \citep[e.g.,][]{bastien1990, fischer1996, Whitney2003, robitaille2006, vandeportal2019, lin2023} have aided the interpretation of the observations connecting polarimetry with the properties and spatial distribution of dust. 

Here, we bring together polarimetry and time-domain analyses to explore a new avenue for detecting unresolved debris disks. Information from different parts of a debris disk are separated in time in a light echo because of light travel effects; the degree of polarization in reflected light also depends on disk location. Lightcurves of different polarized states give unique information not only about local conditions on the disk but the orientation of the unresolved disk relative to the observer. Mapping out a disk in time and polarization also provides needed guardrails for light echo studies. Since light from stellar flares is expected to be unpolarized, the detection of polarized light and its specific lightcurve profile can distinguish a true echo from low-level outbursts from the star following a major outburst.  

The potential hosts of debris disks are plentiful-- approximately 25\% of late-type of main-sequence stars have debris disks. If this frequency hold true for M Dwarfs as well, the echo method could be a valuable technique for constraining debris around these stars. A nice coincidence is that both flare activity and the prevalence of debris from planet formation are higher for younger stars. We expect that potential targets for a echo detection in polarized light are plentiful. With this framing, here next provide a description of our general method for modeling the echo light curves following an impulsive stellar outburst (\S\ref{sec:bg}). Then, in \S\ref{sec:pol}, we introduce polarization in our echo analysis. We summarize our findings, including an assessment of polarization studies for echo detection, in \S\ref{sec:discuss}.

\section{Echo light curves}\label{sec:bg}

The first step in our analysis is to build a framework for predicting the light curve of a source that contains an echo from a debris disk. We envision an impulsive flare at the location of an active star; a distant observer measures the impulse and the subsequent changes in flux as light is reflected by the disk.  

\begin{figure}[!h]
    \centering
    \includegraphics[scale=0.38]{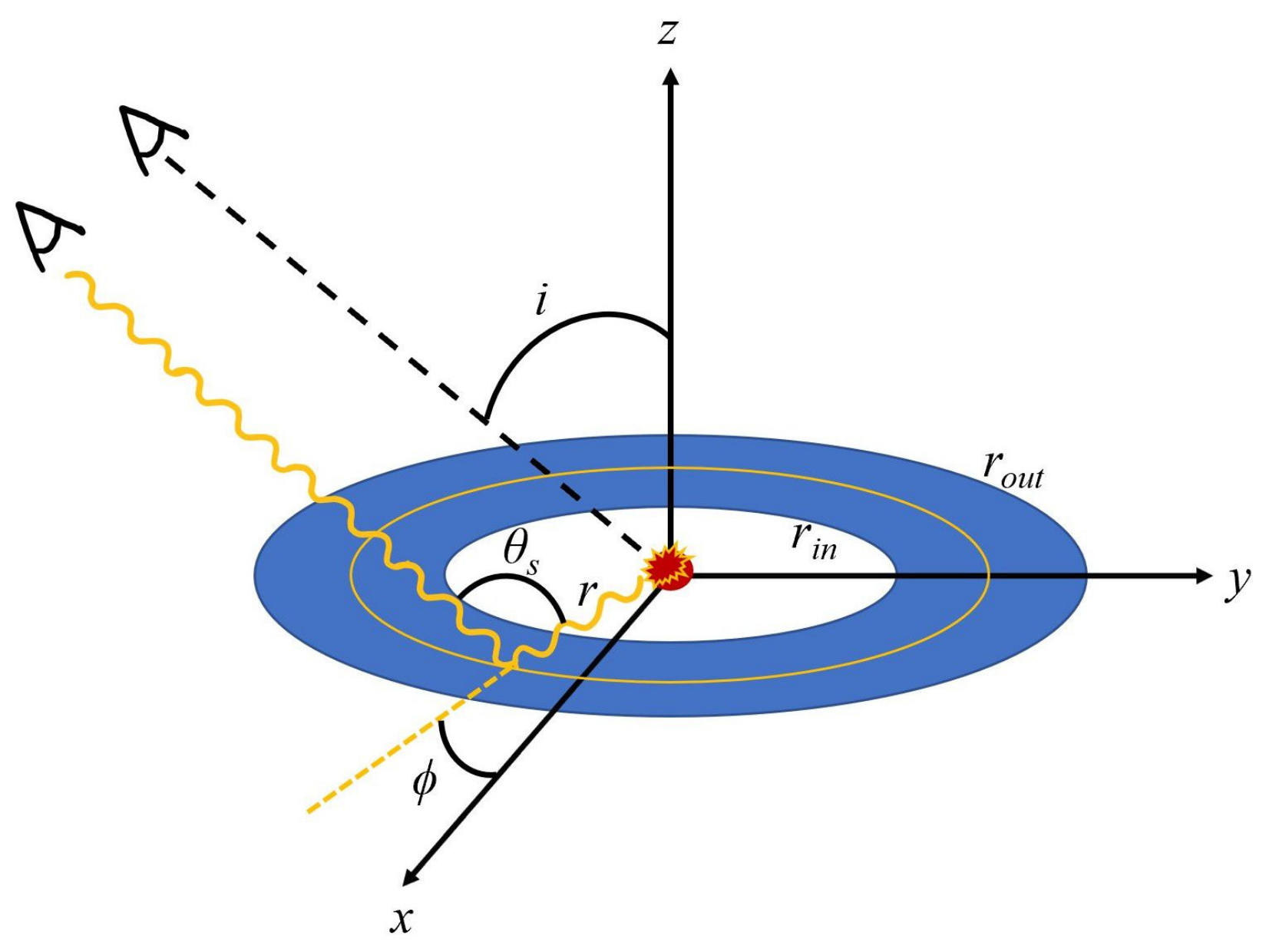}
    \caption{Geometric Disk Model: The debris disk lays flat in the x-y plane with the z axis normal to its surface and surrounds the host star, located at the origin. The yellow ring represents an arbitrary position that flare light had traveled to and reflected off the disk. The path of a single photon is shown by the yellow squiggle, reflecting off of the disk and toward the observer. The angle $i$ is the inclination of the disk as seen by the observer ($0-90^\circ$, which represent face-on and edge-on orientations, respectively). $\phi$ represents the angular position at any point around the disk and $r$ is its corresponding radial position with $r_{in}$ and $r_{out}$ marking the minimum and maximum radial positions. $\theta_{s}$ is the scattering angle for a photon reflected toward the observer. Note: In this orientation, the portion of the disk aligned with the negative y-axis would be defined as the "near-side", as it is inclined toward the observer.}
    \label{fig:geometry}.
\end{figure}

Disk geometry plays a significant role in the shape and timing of the echo profile. Light that scatters off the disk toward the observer will have varying time delays depending on the disk's inner and outer radii and its inclination toward the observer. For a disk in a face-on orientation to the observer, the light scattered from a disk particle at a distance $r$ from the central star will reach the observer at the same post-flare time delay, regardless of its angular position, $\phi$ ($\phi \in [0, 2\pi]$), measured from the positive x-axis of the disk. Thus, $\phi=-\pi/2, \pi/2$ represent the near- and far-sides of the disk, respectively, when the rotation angle of the disk is zero (see Figure~\ref{fig:geometry} for visualization of disk parameters). However, if the disk is inclined, the light scattered from the near-side of the disk will reach the observer first and the light scattered from the far end of the disk will arrive last. This additional time delay helps us to ascertain that a disk is present and constrains the geometric parameters of this disk. For an edge-on disk, there is practically no time delay between the flare itself and the light scattered from the near side of the disk, making it very difficult to differentiate them.

Echo measurements are sensitive to the instrumental setup and how the light curve is sampled. For example, NASA's Transiting Exoplanet Survey Satellite (TESS) mission produced data consisting of 2-minute (short-cadence) and 30-minute (long-cadence) exposures \citep{ricker2015}. Based on light-travel time, the short-cadence data allows us to probe the area within a few AU of the host star, while long-cadence data can be used for areas beyond about 10 AU. For either of these regions, the observed post-flare light curve may contain an echo profile, which can be utilized to determine information about the scattering material that produced it. For the purposes of this paper, we focus on the long-cadence regime.

\subsection{Echo Calculation and Fitting}

To produce model echo light curves, we follow the procedure outlined in \citet{bromley2021}. We assume that the disk is optically and geometrically thin; to a good approximation all of the dust lies in the mid-plane. The formulae used from \citet{bromley2021} are summarized below.

The Draine (\citealt{draine2003}) phase function 
\begin{equation}
    \Phi_\alpha(\theta_s) = 
    \frac{1}{4\pi} 
    \left[\frac{(1-g^2)}{(1+\alpha(1+2g^2)/3}\right] \frac{1+\alpha\cos^2\theta_s}{(1+g^2-2g\cos{\theta_s})^{3/2}},
\end{equation} 
gives the angle-dependent intensity of scattered light,
where $\theta_s$ is the scattering angle, $g$ and $\alpha$ are taken to be 0.429 and 0.114, respectively, in accordance with Draine's parameterization of scattered light with wavelength $0.8~\mu$m from astrophysical dust. 

The scattering angle is
\begin{equation}
    \theta_s = \arccos{\left(-\sin{\phi}\sin{i}\right)}
\end{equation}
where $\phi$ is the angular position of a point on the disk ($\phi = 0$ is aligned with the positive x-axis projected onto the sky) and $i$ is the disk's inclination with respect to the observer ($i=0^\circ= $ face-on, $i=90^\circ= $ edge-on).

The ``brightness'' of any bin of differential area $dA=rdrd\phi$ relative to the incident starlight is
\begin{equation}\label{eq:B}
    B = \frac{C(r) dA}{r^2}\left|1-\phi/\pi\right|\Phi_\alpha(\theta_s)
\end{equation}

The function $C$ carries the dependence of the scattered light on several key physical parameters of the system:
\begin{equation}
    C(r) =\frac{3 Q_\text{eff} M_\text{disk}}{8\pi\rho X r_\text{p,min}(r_\text{out}-r_\text{in})r},
\end{equation}
where $Q_\text{eff}$ is the scattering efficiency of the individual dust grains, taken as $Q_\text{eff}=2$ , consistent with scattering theory for micron-size or larger grains \citep{vandehulst}.
The parameter $X$ is included because we work with average dust properties instead of the underlying size distribution ($n(r_p)$). 
Here, we adopt minimum and maximum particle sizes of $r_{p,min} = 1$~$\mu$m and $r_{p,max} = 1$~mm, respectively, so that $X \approx 30$, where $X$ is the size distribution of the dust particles, and is defined as $X=(\frac{r_{p,max}}{r_{p,min}})^{1/2}$ \citep[cf.]{bromley2021}.

We pair \eqref{eq:B} with the crossing time, defined by \citet{gaidos1994}:
\begin{equation}
    T=\frac{r}{c}\left(1+\sin{\phi}\sin{i}\right).
\end{equation}
We divide up the disk radially (50 bins) and azimuthally (200 bins) into small patches (area $dA$) and assign each patch a characteristic light travel time, $T$, from the above equation. We combine the brightness from all patches that have $T$ falling within some time bin centered at $t_j$ to get the total flux of reflected light in that time bin. Repeating for all patches across the surface of the disk, we generate a full light curve. With this model, an artificial echo can be produced using any desired set of disk parameters. 

To enhance the signal-to-noise, we envision co-adding light curves from multiple flares. Modeling in this manner allows us to treat flares as isotropic and negates the need for a flare position parameter. We use synthetic data with added noise to assess parameter estimation. Using data available from TESS as a guide, we add in noise to simulate observed co-added stellar flare light curves, normalized to a peak flux value of unity. Assuming that the uncertainties in the light curves are Gaussian distributed, we use Python's NumPy package, specifically the \texttt{numpy.random.normal} routine, to generate noise in each flux bin. For this work, we ultimately used a $1\sigma$ uncertainty corresponding to about 10\% that of the TESS data, as noise higher than that often led to degeneracies.

\begin{figure}
    \centering
    \includegraphics[scale=0.8]{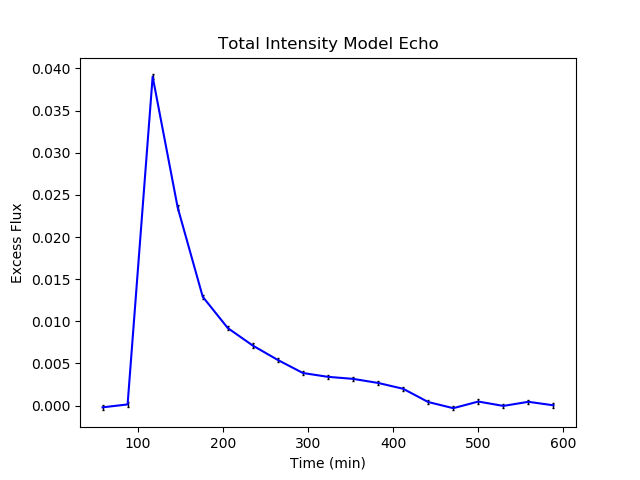}
    \caption{An artificial light echo produced by a model disk with parameters $R_{in} = 50$ AU, $R_{out} = 55$ AU, $i = 45^\circ$, and $M_{disk} = 0.1$ M$_\oplus$, resulting from a flare at time $t=0$~s. The amplitude of the flare itself is 25 times brighter than the flux observed at $t=100$~s  Noise was added on the order of the errors in TESS data for star HD 290527}
    \label{fig:echo1}.
\end{figure}

Figure \ref{fig:echo1} shows an artificial echo for a model debris disk with the following parameters: $R_{in} = 50$ AU, $R_{out} = 55$ AU, $i = 45^\circ$, and $M_{disk} = 0.1$ M$_\oplus$. The vast majority of the scattered flare light comes from the region of the disk nearest to the observer. The flux is brightest here because grains forward scatter light, and contributes to early times in the light curve because of the shorter light travel time between this region and the observer. There is still scattered light from the back end of the disk (visible at a time of around 600 minutes), but it is significantly dimmer. The echo here is shown to peak at about $f = 0.04$, where $f$ is the excess flux above the quiescent starlight. Integrated over the full time sequence, this curve of reprocessed represents an $11\%$ increase over the quiescent starlight over the same duration.

To analyze the ``data'' we have prepared, we fit the data to a disk model with the goal of making parameter estimates and determining whether the light curve is from a debris disk. To achieve this, we pass it through an MCMC fitting algorithm, which carries out a random walk through various disk models to optimize disk parameters and fit the ``data'' light curve. We replicate the model fitting process described in \citet{bromley2021}, with the set-ups and equations summarized here as follows:

We begin with the classic $\chi^2$ measure for determining goodness of fit, defined as
\begin{equation}
    \chi^2 = \sum_{j=1}^{N}\left(F_{data, j} - F_{model}(t_j-t_0, r_{in}, r_{out}, i, M_\text{disk})\right)^2/\sigma_j^2
\end{equation}
where $j$ is the time bin index and $t_0$ represents the time associated with the flare. This fit takes an input data curve and compares it to curves made from random parameters during MCMC fitting to return best-fit disk parameters.

Assuming independent Gaussian errors, the log probability can be defined as 
\begin{equation}\label{eq:logP}
    \ln{\mathcal{P}} = -\frac{1}{2}\chi^2 -\frac{1}{2}\sum_{i}\ln\left({2\pi\sigma_i^2}\right) + \ln{\mathcal{P}_{priors}}
\end{equation}
where $\mathcal{P}_{priors}$ is defined as
\begin{equation}
    {\cal P}_\text{priors} \sim  
    \begin{cases}
    \ \sin(\inc) \exp(-\mfit/M_\tau) & \ \ 0^\circ \leq \inc \leq 90^\circ,
    \ 0 \leq \mfit/\Mearth, \text{and} \\
    & r_{min} \leq r_{in} < r_{out} \leq r_{max}, 
    \\
    0 & \text{otherwise.}
    \end{cases} 
\end{equation}
Equation (\ref{eq:logP}) corrects a typographical error in
\citet{bromley2021}.
In the above equation, $M_\tau$ refers to a scaled mass above which the optically thin assumption for our disk becomes invalid. This condition is 
\begin{equation}
    M_\tau \approx 0.36\left[\frac{r_{in}}{30 AU}\right]
    \left[\frac{r_{out}}{60 \text{au}}\right]
    \left[\frac{\rho}{2\ \text{g/cm}^3}\right]
    \left[\frac{X}{31.6}\right]
    \left[\frac{r_{p,min}}{1\ \mu \text{m}}\right] M_\oplus .
\end{equation}

\begin{figure}
    \centering
    \includegraphics[scale=0.6]{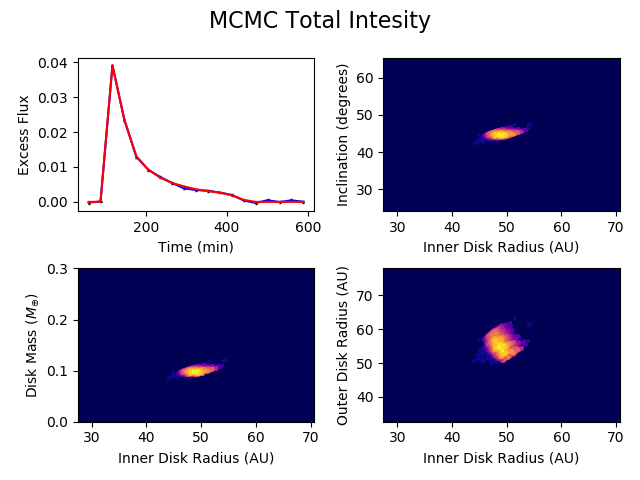}
    \caption{Top left: the blue line represents artificial echo data; the red line is the echo profile predicted by best-fit parameters. Top right: inner disk radius vs inclination. Bottom left: inner disk radius vs disk mass. Bottom right: inner disk radius vs outer disk radius. As we can see in the top right and both bottom parameter maps, the MCMC fitting process yields results that match well to the input artificial data (same disk parameters were used as those in Figure 2). Noise was added to the light curve on the order of the errors in TESS data for star HD 290527. }
    \label{fig:mcmc1}
\end{figure}

With the likelihood parameters in place, we utilize the emcee Markov-chain Monte Carlo (MCMC) package \citep{emcee2012} to map out the likelihood in the space of possible disk parameters. The created sample allows for estimation of best-fit values and confidence regions in parameter space. Figure~\ref{fig:mcmc1} shows the result of this MCMC process with a generated ``data'' echo of the same parameters shown in Figure~\ref{fig:echo1}. This summed echo signal has had noise added on the order of $10\%$ of typical errors observed in 10th magnitude flaring TESS stars to resemble real data. The resulting map of likelihood distribution in parameter space shows that it is feasible to extract disk parameters from an echo light-curve.

While the MCMC fit of total intensity shows promising results, we face a difficulty when approaching edge-on orientations. The simulations in this model may still show successful fits (see left 4 panels of Figure \ref{fig: fit_results_edge}),but distinguishing this portion of the light from post-flare data would be difficult. Light curves of stellar flares may have unexpected post-peak structure or subflaring that could appear similar to an echo. Next, we consider the assistance that polarized scattered light could provide in disentangling scattered light and additional fitting opportunities it allows.

\section{Polarized echo light curves}\label{sec:pol}
Light scattered by circumstellar dust will be polarized with respect to the flare that originally produced it. Polarization is a powerful tool. Detecting excess polarization post-flare enables clear identification of an echo as opposed to a second, smaller flare. Analyzing the polarized signal may yield the origin of the echo and the architecture of the scattering surface.  As in \citet{kawate2019, bianda2005}, we assume that the initial flare is fully unpolarized and use the formalism of \citet{yangli2019} to make polarization calculations.

\subsection{Polarization Calculations}
In their \citeyear{yangli2019}
paper, Yang and Li laid the groundwork for calculation of polarization for a debris-disk geometry. They defined the following trigonometric functions needed to make calculations of the Stokes parameters $Q$ and $U$: 

\begin{equation}
    \cos{\phi'(\phi)}=-\frac{\sin{\phi}}{\sqrt{\cos^2{i}\cos^2{\phi}+\sin^2{\phi}}}
\end{equation}
\begin{equation}
    \sin{\phi'(\phi)}=\frac{\cos{i}\cos{\phi}}{\sqrt{\cos^2{i}\cos^2{\phi}+\sin^2{\phi}}}
\end{equation}

These trigonometric functions depend on $\phi'$, the projected angular measurement on the sky, $\phi$, the physical angular measurement on the disk, and $i$, the inclination of the disk. We use $\cos{(2\phi'(\phi))}$ and $\sin{(2\phi'(\phi))}$ due to the translation of the Stokes vector from the scattering frame to the lab frame.

Within this setup, the expressions for Stokes $I, U,$ and $Q$  are:
\begin{equation}
    I \propto \int_0^{2\pi}Z_{11}(\theta_s(\phi))d\phi
\end{equation}
\begin{equation}
    Q \propto \int_0^{2\pi}Z_{21}(\theta_s(\phi))\cos{(2\phi'(\phi))}d\phi
\end{equation}
\begin{equation}
    U \propto \int_0^{2\pi}Z_{31}(\theta_s(\phi))\sin{(2\phi'(\phi))}d\phi
\end{equation}
where $Z_{ij}$ represent terms of the Mueller matrix, which is responsible for determining how polarization states change upon scattering. We calculated these terms for Mie scattering using the pySCATMECH \texttt{Free\_Space\_Scatterer} module. For this paper, we set the parameters of the pySCATMECH module to the following: $\lambda=800\ \text{nm}, n_1 = 1, n_2 = 1.33, r=1\ \mu\text{m}$, where $\lambda$ is the wavelength of observed light, $n_1$ and $n_2$, are the indices of refraction for the medium and scattering material (set as 1 for vacuum and 1.33 for ice), respectively, and $r$ is the radius of the scattering particles. In the Stokes formalism, I denotes total intensity, Q represents the vertically and horizontally polarized components of the light, and U represents the $\pm 45^\circ$ polarized light.

This formalism is constructed by taking the rotation orientation of the disk as $0^\circ$, meaning that the ``bottom'' of the disk is inclined toward the observer. The ``right'' side of the disk inclined toward the observer would correspond to a rotation orientation of $90^\circ$, and so forth. 
Due to the symmetry of the system in the $0^\circ$ rotation case, as in \citet{yangli2019}, light from portions of the disk that are oppositely polarized in Stokes $U$ reach the observer at the same time, so Stokes $U$ cancels out to zero. The polarization fraction in this model can be defined relatively simply as
\begin{equation}
    p_s = \frac{Q}{I} = \frac{\int_0^{2\pi}Z_{21}(\theta_s(\phi))\cos{(2\phi'(\phi))}d\phi}
    {\int_0^{2\pi}Z_{11}(\theta_s(\phi))d\phi}
    \label{eq: p-s}
\end{equation}
where $p_s > 0$ describes horizontal polarization and $p_s < 0$ describes vertical polarization.

While eq.~\ref{eq: p-s} is very useful when Stokes U = 0, this case holds only for specific orientations where oppositely polarized light from different sides of the disk reaches the observer at the same time, leading to zero U. Because a debris disk can be oriented to any rotation angle relative to the observer, we utilize the rotation angle as a parameter and not keep it fixed at zero. Thus, we adjust the formulas to be more flexible to changes in rotation and to calculate the total polarization fraction due to contributions from both Stokes $Q$ and $U$. The total polarization fraction is then defined as
\begin{equation}
    p_s = \frac{Q^2 + U^2}{I^2}
\end{equation}
This scattering will not result in 100\% polarization of light, so $I^2 > Q^2 + U^2$, and thus $p_s < 1$.

\subsection{Polarization Examples}

We now calculate the polarization fraction at any point from a model debris disk and then predict the polarized light curve. After calculating an unpolarized light curve as in Figure 2, we map the derived polarization fractions for each disk element to its corresponding place on the unpolarized light curve. Figure 4 shows the light curves in total intensity, Stokes $Q$, Stokes $U$, and the summed polarization where $I \geq \sqrt{Q^2 + U^2}$. 
These are predicted for a system with the same parameters as shown in Figure 2 with the addition of the rotation parameter, which was set at $10^\circ$. A $10^\circ$ rotation means that the portion of the disk inclined nearest to the observer is rotated counter-clockwise by 10 degrees relative to the negative y axis projected onto the sky.

\begin{figure}[!h]
    \centering
    \includegraphics[scale=0.6]{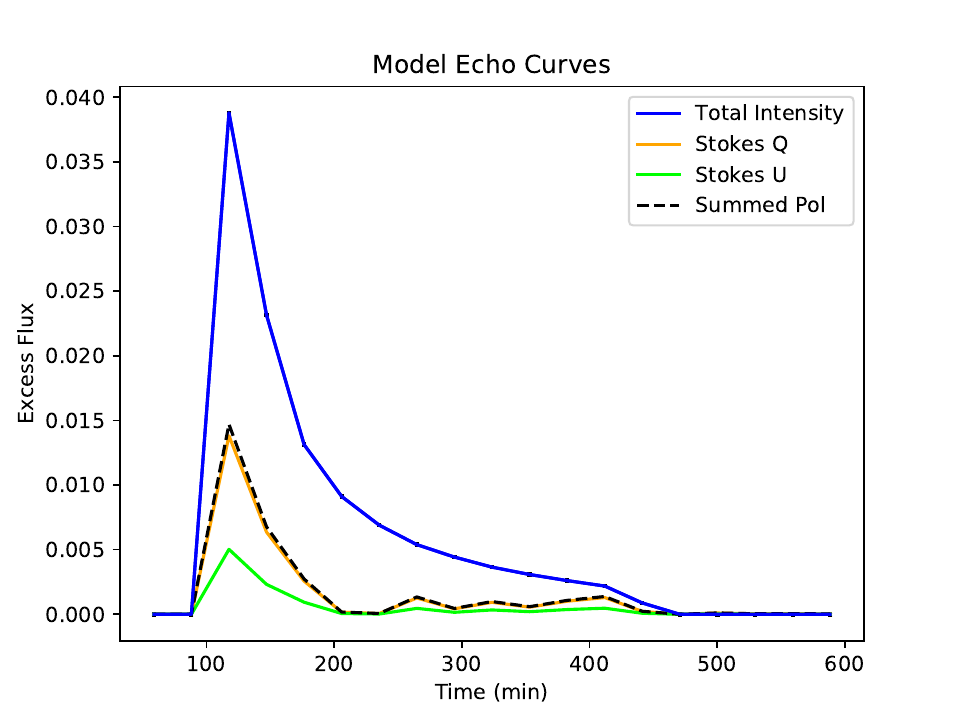}
    \caption{A set of model echo light curves without noise: These curves are calculated from the same parameters used in figure 2, with the addition of a rotation parameter for polarization, which is set to $10^{\circ}$. The blue line depicts the total intensity of the echo light, while orange and lime show stokes Q and U, respectively. The black dashed line is the total polarization curve, $\sqrt{Q^2+U^2}$}
    \label{fig:no_noise}
\end{figure}

Noise for the polarized light curved was approximated by combining the method used for creating randomly generated noise from the unpolarized case with potential interstellar medium (ISM) scattering. For heavily extincted stars, the intrinsic polarization of the star may exceed the polarization signal from any echo \citep{angarita2023}. It is then essential to have very high signal to noise data to separate the echo signal from the stellar signal. However, for nearby bright stars, inclusion of ISM polarization may not be significant compared to the polarized light echo method. The maximum percentage of photons polarized can be defined as $P_\text{max}/A_V \leq 3 \text{ percent per magnitude of extinction}$ \citep{Das}. Pairing this with the reported average of 0.2 magnitudes of extinction for a survey of 9000 flare stars in the galactic halo \citep{Gontcharov}, a maximum of 0.6\% of photons would be polarized by the ISM when our primary targets are flare stars at high galactic latitudes. However, for regions of lower latitudes with higher density of dust in the ISM, this can begin to pose a problem for our polarization measurements. 

To test the ability of the polarized light echo method with significant polarization due to the ISM, we consider HD 290527, a flaring A-type star. Using TESS data and the NASA/IPAC archive to find an extinction value of $A_V = 1.03 \text{ mag}$, we add noise to our artificial light curves via the Poisson noise produced from the excess polarized photons. With this factored in to our original noise estimates, we generate our artificial light curve data.

\begin{figure}[!h]
    \centering
    \includegraphics[width=.48\textwidth, height=.31\textwidth]{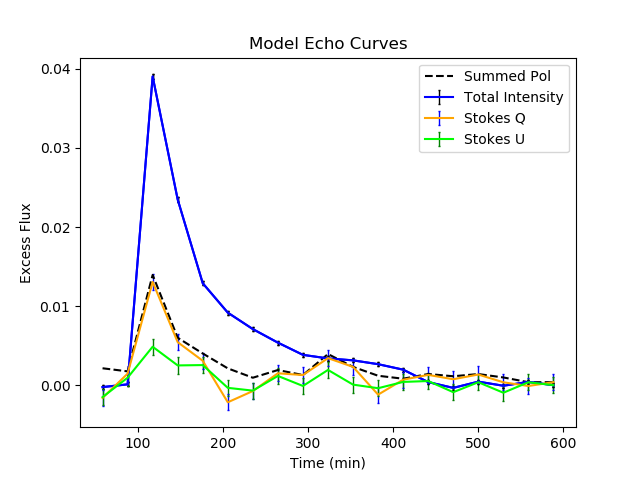}
    \caption{A set of model echo light curves with noise: These curves are calculated from the same parameters used in figure 2, with the addition of a rotation parameter for polarization, which is set to $10^{\circ}$. The blue line depicts the total intensity of the echo light, while orange and lime show stokes Q and U, respectively. The black dashed line is the total polarization curve, $\sqrt{Q^2+U^2}$. Noise was added as described in the text.}
    \label{fig:pol_curve}
\end{figure}

As shown in Figure \ref{fig:pol_curve} 
there is a peak in flux in all three light curves from the portion of the disk nearest to the observer. This produces the most scattered light from our perspective due to the forward scattering properties of optically thin disks \citep[]{draine2003}.

In this orientation, the peak brightness observed in Stokes $Q$ is +Q, which is horizontally polarized. As time passes and we receive signal from other portions of the disk, we observe more -Q (vertical polarization) which cancels out much of the +Q light until we get virtually no signal after about 200 minutes.

Comparing relative fluxes as we did with Figure \ref{fig:echo1}, the total intensity curve from Figure \ref{fig:pol_curve} still reaches $11\%$ of the quiescent starlight. The summed polarization from the same set of curves reaches $45\%$ of the total intensity echo, totaling $5\%$ of the quiescent starlight. This presents a feasible value for observation and would allow observers to confirm the occurrence of scattered light in order to disentangle it from the flare lightcurve. 

\begin{figure}[!ht]
    \centering
    \includegraphics[width=.48\textwidth, height=.31\textwidth]{error_curves45inc.png}
    \includegraphics[width=.48\textwidth, height=.31\textwidth]{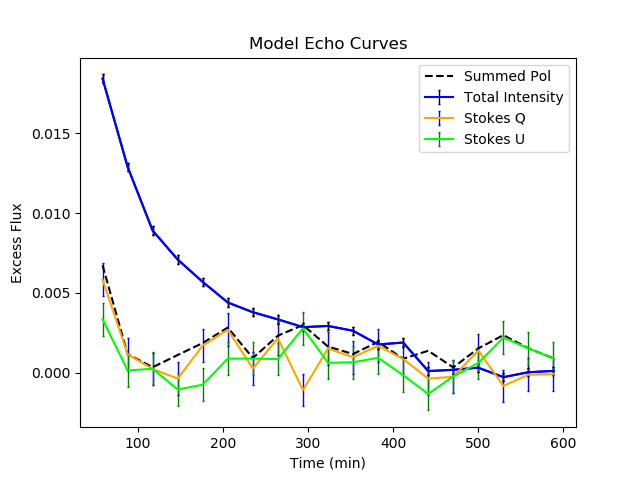}
    \caption{Light Echo Plots: In each of the above plots, the blue lines represent total intensity of the light echo, the orange lines represent light polarized in Stokes Q, and the magenta lines represent light polarized in Stokes U. The black dashed line shows the summed polarization $I_{pol} = \sqrt{Q^2+U^2}$. The left figure shows light curves for the same parameters shown in Figure 2. The right figure shows light curves for the same parameters shown in Figure 2 except for inclination, which has been changed to $85^\circ$ to show an edge-on orientation. Both curves have had noise added on the order of a TESS flare star with 64 flare events.}
    \label{fig:curves}
\end{figure}

In Figure \ref{fig:curves}, we compare the same light curves from Figure \ref{fig:pol_curve} with curves for a model disk inclined at $i=85^\circ$. In both cases, the first peak of each light curve coincides. The most notable change is the peak in Stokes Q between 350 and 400 minutes. This is the most useful light from this set of curves, as it will be most easily distinguishable from the flare light itself. The peaks from the front end of the disk in this edge-on case will likely be too overpowered by flare light to be easily extracted from data, making the latter portion of the light curve extremely valuable. 

\begin{figure}[!ht]
    \centering
    \includegraphics[width=.49\textwidth]{TI_mcmc45inc.png}
    \includegraphics[width=.49\textwidth]{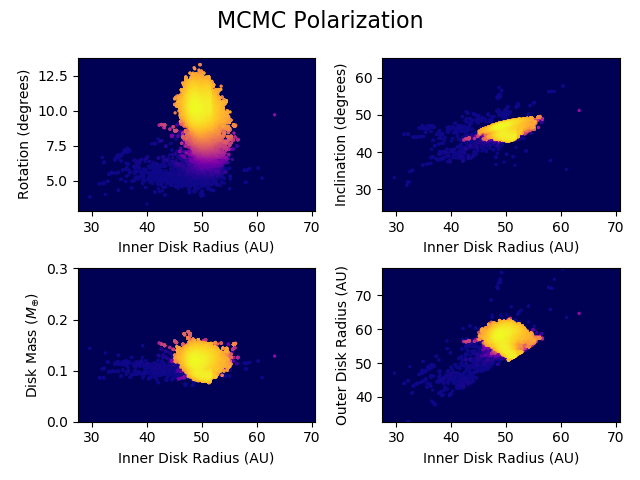}
    \caption{MCMC fit results in parameter space for the unpolarized (left 4 panels) and polarized (right 4 panels) echo curves shown in the left panel of Figure \ref{fig:curves}.}
    \label{fig: fit_results}
\end{figure}

\begin{figure}[!ht]
    \centering
    \includegraphics[width=.49\textwidth]{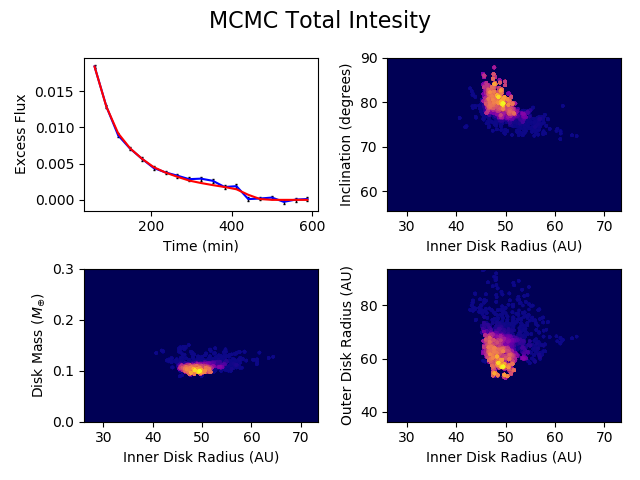}
    \includegraphics[width=.49\textwidth]{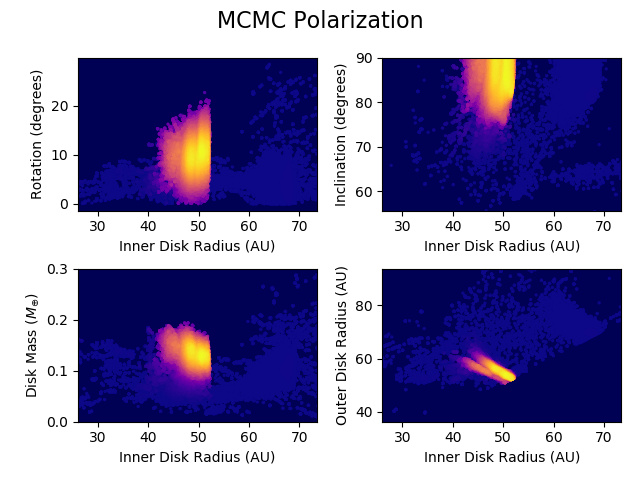}
    \caption{MCMC fit results in parameter space for the unpolarized (left 4 panels) and polarized (right 4 panels) echo curves shown in the right panel of Figure \ref{fig:curves}.}
    \label{fig: fit_results_edge}
\end{figure}

To showcase the fitting capabilities of the polarized model compared to total intensity, we run the same MCMC algorithm as discussed above on the total intensity of light and both Stokes $Q$ and $U$ for the cases of $45^\circ$ and $85^\circ$ inclination. The results are shown in \ref{fig: fit_results} ($45^\circ$) and \ref{fig: fit_results_edge} ($85^\circ$) and are given in Table 1.

\begin{figure}[!ht]
    \centering
    \includegraphics[width=.49\textwidth]{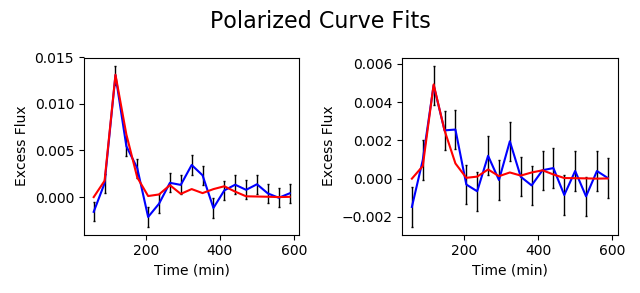}
    \includegraphics[width=.49\textwidth]{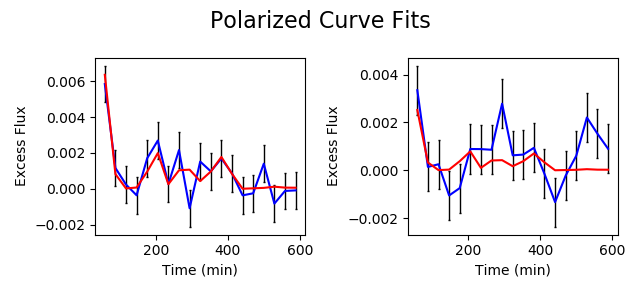}
    \caption{The two panels on the left show the Stokes Q and U fit results, respectively, at 45$^\circ$ inclination. The two panels on the right show the same fits for $85^\circ$ inclination. The first 3 panels show strong curve-fitting. The fourth panel does not converge as cleanly, but it also has a signal strength of about an order of magnitude weaker than the others, and is likely inconsequential due to this.}
    \label{fig: pol_fits}
\end{figure}

\begin{deluxetable}{cccccccccccc}[!h]
\tabletypesize{\footnotesize}
\tablecolumns{12}
\tablewidth{0pt}
\tablecaption{Total Intensity vs Polarization}\label{tab:int_v_pol}
\tablehead{\colhead{Observed Light} & \colhead{$r_{in}$} & \colhead{$r_{in}$ $95\%$ range} & \colhead{$r_{out}$} & \colhead{$r_{out}$ $95\%$ range} & \colhead{inc} & \colhead{inc $95\%$ range}  & \colhead{mass} & \colhead{mass $95\%$ range} & \colhead{rot} & \colhead{rot $95\%$ range}
\vspace{-0.2cm}
\\
 & \colhead{(AU)} & \colhead{(AU)} & \colhead{(AU)} & \colhead{(AU)} & \colhead{(deg)} & \colhead{(deg)} & \colhead{\Mearth} & \colhead{\Mearth} & \colhead{(deg)} & \colhead{(deg)}}
\decimals
\startdata
T.I. 1 & 48.60 & 47.47$-$50.66 & 54.64 & 52.66$-$57.98 & 44.57 & 44.22$-$45.39 & 0.10 & 0.09$-$0.10 & NA & NA \\
Pol 1 & 48.77 & 48.12$-$51.19 & 57.95 & 52.58$-$59.73 & 46.83 & 44.14$-$47.55 & 0.12 & 0.09$-$0.13 & 10.29 & 9.22$-$11.15 \\
Run 1 & 50 & & 55 & & 45 & & 0.1 & & 10 & \\
T.I. 2 & 49.52 & 46.03$-$53.49 & 56.98 & 56.32$-$73.68 & 79.7 & 75.57$-$82.44 & 0.10 & 0.10$-$0.12 & NA & NA \\
Pol 2 & 50.45 & 41.93$-$68.30 & 53.99 & 52.90$-$121.47 & 86.25 & 27.35$-$89.48 & 0.13 & 0.07$-$0.15 & 10.81 & 1.74$-$13.51 \\
Run 2 & 50 & & 55 & & 85 & & 0.1 & & 10 & 
\enddata
\tablecomments{T.I. and Pol represent total light intensity and polarization, respectively. T.I. 1 indicates the results from the MCMC fit in total intensity at $i=30^\circ$ and T.I. 2 indicates the results at $i=85^\circ$ (the same is true for Pol 1 and Pol 2, respectively). The rows labeled Run 1 and Run 2 show the input values for the disk parameters in each fit process.
}
\end{deluxetable}

While the results for the polarization fit may not look as accurate as the total intensity case, it is important to note that in both cases the intensity of the polarized curves is approximately an order of magnitude weaker, but the polarization fit is still able to return correct parameter values in a competitive way. On top of this, the edge-on case for the total intensity light curve cannot be trusted to return these results in practice, as we may not be able to accurately extract this portion of the light curve from the flare light. So, while the polarized fit in the edge-on case does look a little messier, it would actually be more accurate than the total intensity in practice.

\subsection{Face-on Orientation}

A prediction of the model for face-on orientations is that light from any angular position on the disk (at the same radial position) would reach the observer at the same time; any light that has been polarized vertically will be accompanied to the same degree by light that has been polarized horizontally and vice-versa, and the same happens for $\pm 45^\circ$ polarized light. Polarization then vanishes and we are left with total intensity of light. This is an acceptable limitation, however, due to the knowledge that a purely face-on orientation is unlikely and the face-on orientation does not come with the problem of echo and flare photons arriving at the same time.

\section{Summary and Next Steps}\label{sec:discuss}

Light echoes offer the promise of detecting protoplanetary disks in the time domain. Here, we consider an optically and geometrically thin disk model with calculated polarization states of scattered photons to predict the structure of post-flare light echoes. Using this model, we feed artificial echo data to a Markov-Chain Monte Carlo fitting process. We find encouraging results for the possibility of parameter recovery, which leads us to believe that from polarized echo data, we should be able to not only uncover the presence of circumstellar material, but determine its geometrical shape, orientation, and mass content as well. 

While previous work such as \citet{bromley2021} has presented a similar outcome using only unpolarized light, this work demonstrates that the inclusion of polarization in the model poses a strong asset to its ability to return correct disk parameters. This is made clear by Table 1, which shows competitive results in the $95\%$ confidence range for each disk parameter in our model, even with total intensity peak fluxes around 3-4 times higher than that of polarization fluxes. The primary reasons that the polarization model is advantageous are:

\begin{enumerate}[topsep=2pt, partopsep=2pt, itemsep=10pt,parsep=2pt]
    \item flare light will likely be unpolarized \citep{bianda2005, kawate2019}.
    \item A predictable relationship exists between the polarized components in the light echoes from a dusty disk;
    \item There is a guaranteed time delay between flare and echo light in at least one polarized component, even for edge-in disks.
    \item Dual fitting in both Stokes $Q$ and $U$ results in better parameter constraints at a lower intensity than modeling with the total intensity alone.
    \item Polarized light can be used to confirm if post-flare structure is due to scattering (echoes) or can be attributed to subflaring.
\end{enumerate}

Throughout this study, we neglected to vary the size distribution parameter, $X$ (introduced in Eq.~(4)), which sets the total disk mass and its scattering potential. Specifically, we fixed the scattering radius at the minimum particle size of $1 \mu m$ as the smaller particles in the distribution will be responsible for the majority of scattering. Grains smaller (larger) than 1~$\mu$m typically produce a stronger (weaker) polarization signature. While deriving a best-fit minimum particle size during the MCMC process is possible, this approach is computationally expensive and is left for future work.

The promise of incorporating polarization into light-echo analyses has significant challenges. As with unpolarized light-echo detection of debris disks, echoes are faint compared with the quiescent flux of the stellar host. Multiple flares are required to build up a statistically significant light-echo signal in a composite post-flare light curve. An instrument such as the Zwicky Transient Facility \citep[e.g.,]{Bellm19}, equipped with a polarimeter for measurements in Stokes $Q$ and $U$, could provide high-quality data for echo detection in polarized light. 
Long-term monitoring of flare stars via a dedicated CubeSat \citep[e.g.,][]{shkolnik2018} with polarization capabilities \citep{dicandia2020} would allow similar time coverage. These possibilities would enable more precise measurement of the debris disk frequency among late M-type stars and better characterization of the geometry and grain properties of M dwarf debris disks.

\section{Acknowledgements}
We thank an anonymous referee for thoughtful suggestions to improve this paper.

\pagebreak
\bibliography{bib}{}

\end{document}